\documentstyle[12pt, aaspp4]{article}

\def\ltsima{$\; \buildrel < \over \sim \;$}
\def\simlt{\lower.5ex\hbox{\ltsima}}
\def\gtsima{$\; \buildrel > \over \sim \;$}
\def\simgt{\lower.5ex\hbox{\gtsima}}

\begin{document}

\title{An Accretion Model for Anomalous X-Ray Pulsars}

\author{Pinaki Chatterjee$^1$, Lars Hernquist$^2$ \& Ramesh Narayan$^3$}
\affil{Harvard-Smithsonian Center for Astrophysics, 60 Garden
Street, Cambridge, MA 02138}
\footnotetext[1]
{pchatterjee@cfa.harvard.edu}
\footnotetext[2]
{lars@cfa.harvard.edu}
\footnotetext[3]
{rnarayan@cfa.harvard.edu}

\medskip

\begin{abstract}

We present a model for the anomalous X-ray pulsars (AXPs) in which the
emission is powered by accretion from a fossil disk, established from
matter falling back onto the neutron star following its birth.  The
time-dependent accretion drives the neutron star towards a
``tracking'' solution in which the rotation period of the star
increases slowly, in tandem with the declining accretion rate.  For
appropriate choices of disk mass, neutron star magnetic field strength
and initial spin period, we demonstrate that a rapidly rotating
neutron star can be spun down to periods characteristic of AXPs on
timescales comparable to the estimated ages of these sources.  In
other cases, accretion onto the neutron star switches off after a
short time, and the star becomes an ordinary radio pulsar.  Thus, in
our picture, radio pulsars and AXPs are drawn from the same underlying
population, in contrast to models involving neutron stars with
ultrastrong magnetic fields, which require a new population of stars 
with very different properties.

\end{abstract}

\keywords{stars: neutron -- pulsars: general -- accretion, accretion disks 
-- X-rays: stars}

\section{Introduction}

Anomalous X-ray pulsars (AXPs) are a subclass of X-ray pulsars
comprising approximately half a dozen members with certain
well-determined properties (e.g. Mereghetti \& Stella 1995; van
Paradijs, Taam \& van den Heuvel 1995).  AXPs are sources of pulsed
X-ray emission with steadily increasing periods lying in a narrow
range, $P\sim 6 - 12$ seconds, and having characteristic ages
$P/2\dot{P} \sim 10^{3} - 10^{5}$ years.  While it is generally
believed that AXPs are neutron stars, they differ significantly from
other manifestations of these objects.  AXPs have periods more than an
order of magnitude larger than those of typical radio pulsars.
Compared with high-mass X-ray binary pulsars, AXPs have relatively low
luminosities, $L_x \sim 10^{35} - 10^{36}$ erg/sec, and soft spectra.
Generally, AXP spectra are well-fitted by a combination of blackbody
and power-law contributions, with effective temperatures and photon
indices in the range $T_e \sim 0.3 - 0.4$ keV and $\Gamma \sim 3 - 4$,
respectively.  In addition, AXPs have no detectable binary companions
and at least two have been associated with young supernova remnants.

A particularly uncertain aspect of AXPs is the energy source that
supports their X-ray emission.  From timing measurements, it is clear
that they cannot be rotation-powered.  For values characteristic of
AXPs, the rate of loss of rotational energy is $|\dot{E}| \equiv
4\pi^2 I \dot{P} /P^3 \approx 10^{32.5}$ erg/sec, orders of magnitude
smaller than their X-ray luminosities.  Motivated by this discrepancy,
two competing theories for AXPs have emerged.  In one, the energy
source for the X-rays is internal and the AXPs are modeled as
isolated, ultramagnetized neutron stars (``magnetars''; Duncan \&
Thompson 1992), powered either by residual thermal energy (Heyl \&
Hernquist 1997a) or by magnetic field decay (Thompson \& Duncan 1996).
The observed X-ray luminosities place severe limitations on both
possibilities.  If residual thermal energy drives the emission, then
the envelope of the neutron star must consist primarily of light
elements (Heyl \& Hernquist 1997b).  If magnetic field decay supplies
the energy, then non-standard decay processes may be required unless
the field is $B\simgt 10^{16}$ G (Heyl \& Kulkarni 1998).  In either
case, the measured periods and estimated period derivatives and ages
are consistent with field strengths $B\sim 10^{14} - 10^{15}$ G.
These values are similar to those inferred for soft gamma repeaters
(SGRs) from timing data (e.g. Kouveliotou et al.  1998, 1999; but see,
however, Marsden, Rothschild \& Lingenfelter 1999).

The second class of theories for AXPs invokes accretion to power the
X-ray emission.  Several variants of this hypothesis have been put
forward.  Mereghetti \& Stella (1995) suggested that AXPs are neutron
stars with magnetic fields similar to those of radio pulsars,
accreting from binary companions of very low mass.  Wang (1997)
proposed that the AXP candidate RX J0720.4-3125 is an old neutron star
accreting from the interstellar medium.  (For an interpretation of
this object as an ultramagnetized star, see Heyl \& Hernquist 1998a.)
In another model, AXPs are descendants of high-mass X-ray binaries and
are accreting from the debris of a disrupted binary companion,
following a period of common-envelope evolution (van Paradijs et al.
1995; Ghosh, Angelini \& White 1997).  In all these analyses, the
magnetic fields inferred from the observed luminosities and spin
periods is $B \simlt 10^{12}$ G, if the neutron star is accreting near
its equilibrium spin period.

A potential advantage of accretion-powered models for AXPs over those
based on ultrastrong magnetic fields is that these neutron stars then
have intrinsic properties similar to those of ordinary radio pulsars
and luminous X-ray pulsars.  However, efforts to detect binary
companions of AXPs have been unsuccessful and have placed severe
limits on companion masses (e.g. Mereghetti, Israel \& Stella 1998; 
Wilson et al. 1998).  Moreover, it is clear that if AXPs are
accreting, then they cannot be in equilibrium with their accretion 
disks, since all these objects are observed to spin down.  Ghosh et 
al. (1997) suggest that this deficiency can be overcome by {\it 
time-dependent} accretion, but do not demonstrate if a young, rapidly 
spinning neutron star can be spun down to periods $P\sim 10$ seconds 
on timescales consistent with, e.g., age estimates based on 
associations with supernova remnants (e.g. Vasisht \& Gotthelf 1997).

In this paper, we propose that AXPs are indeed accreting, but that the
accretion disk was established as part of the process which formed the
neutron star, through fallback of material from the progenitor
following the supernova explosion.  After an initial phase of
transient evolution, the accretion rate in this scenario declines
steadily as the material in the disk is depleted.  At late times, the
accretion rate evolves self-similarly, and the torques on the star
spin it down in a regular manner.  We demonstrate that for plausible
choices of the parameters, this time-dependent accretion yields
periods, ages, and luminosities consistent with the observed
properties of AXPs.

\section{Accretion from a Debris Disk}

Following the formation of a neutron star through the core collapse of
a massive progenitor, it is plausible that mass ejection during the
explosion will not be perfectly efficient and that a small amount of
mass, $M_{fb}$, can fall back.  The amount of this fallback is
uncertain, but various lines of argument suggest that $M_{fb} \simlt
0.1 M_{\odot}$ is not unreasonable (Lin, Woosley \& Bodenheimer 1991;
Chevalier 1989).  Much of this material will be directly accreted by
the neutron star, and some may be expelled when the accretion rate
exceeds the Eddington limit, depending on the relative contribution of
neutrino emission to the energy loss rate.  However, some of the
fallback is likely to carry sufficient angular momentum that it will
settle into an accretion disk of mass $M_{d}$ around the neutron star.
The relationship between $M_{fb}$ and $M_{d}$ depends on the detailed
properties of the progenitor, but our model for AXP formation does not
require large disk masses and can, in principle, operate even if
$M_{d} \ll M_{fb}$.

The fate of the material settling into the disk is uncertain, but the
problem we are describing is similar to tidal disruption of stars by
massive black holes.  Numerical simulations show that material
captured by the black hole in such an event will circularize into an
accretion disk on roughly the local dynamical time and that the
subsequent evolution of the disk through viscous processes can be
characterized in a simple manner (Cannizzo, Lee \& Goodman 1990).  In
our analysis, we assume that fall back will establish a thin accretion
disk on a timescale $T \approx 1$ msec.  Detailed calculations show
that the long-term behavior of the system is insensitive to the
numerical value of $T$.

The accretion disk will exist for radii only beyond the magnetospheric
radius $R_{m} \approx 0.5 r_{A} = 6.6 \times 10^{7} B_{12}^{4/7}
\dot{m}^{-2/7}$ cm (Frank, King \& Raine 1992), where the Alfv\'en
radius, $r_{A}$, indicates the transition point at which the flow
becomes dominated by the magnetic field, and $B_{12}$ is the surface
magnetic field strength of the star in units of $10^{12}$ G.  We
parameterize the mass accretion rate through the disk, $\dot{M}$, in
ratio to the Eddington rate, $\dot{M}_{E}=9.75 \times 10^{17}$ g/s, by
$\dot{m} \equiv \dot{M}/\dot{M}_{E}$.  (In numerical estimates, we
always take the neutron star mass and radius to be $M_{*}=1.4
M_{\odot}$ and $R_{*}=10^6$ cm, respectively.)  Note, however, that we
allow for the possibility that the mass accretion rate onto the
surface of the star, $\dot{M_X}$, will be different from $\dot{M}$, if
some of the material is driven from the system prior to reaching the
neutron star surface.  Thus, the X-ray luminosity will be set by
$\dot{M_X}$ rather than by $\dot{M}$, but we will generally take the
location of $R_m$ and the torque applied to the neutron star to be
determined by $\dot{M}$.

Whether or not the disk will influence the spindown of the neutron
star or lead to accretion-induced X-ray emission will depend on the
location of $R_{m}$ relative to the light cylinder radius,
$R_{lc}=c/\Omega$, and the corotation radius $R_c$ defined by
$\Omega=\Omega_K(R_c)$, where $\Omega$ is the angular rotation
frequency of the star and $\Omega _K(R)$ is the Keplerian rotation
rate at radius $R$ (e.g. Shapiro \& Teukolsky 1983).  If at any time
$R_{m} > R_{lc}$, we assume that the disk will effectively evolve
independently and will not affect the neutron star.  Thus, we
anticipate that even if all neutron stars acquire debris disks through
fallback, most will live as ordinary radio pulsars, for appropriate
choices of $M_{d}$, $B$, and initial spin period, $P_0$.  In other
cases, accretion will be permitted and the neutron star will behave as
an accreting X-ray source.  As we shall see, according to this model
it is possible for a given neutron star to spend a portion of its life
either as a radio pulsar or as an accreting source.

During accretion, the interaction between the disk and the star can
lead to several distinct phases of evolution.  If $R_m \gg R_c$, which
corresponds to $\Omega \gg \Omega_K(R_m)$, the ``propeller'' effect
will operate, so that accretion onto the surface of the star will be
inefficient owing to centrifugal forces acting on the matter
(Illarionov \& Sunyaev 1975).  The details of this process are
uncertain.  During the propeller phase, we assume that $\dot{M_X} \ll
\dot{M}$ but that spindown will nevertheless be efficient.  In other
words, matter will be driven to $R_m$ by viscous processes where it
can torque the star, but mass ejection occurs subsequently before the
material reaches the star's surface.  Thus, such an object will spin
down rapidly but will be X-ray faint.  Later, as $\Omega$ approaches
$\Omega _K(R_m)$, the system enters a quasi-equilibrium ``tracking''
phase, in which the spin of the star roughly matches the rotation
period of the disk at $R_m$.  However, because the mass accretion rate
declines steadily, a true equilibrium is never attained, unlike in the
case of accreting neutron stars in binaries.  In this phase, we assume
$\dot{M_X} \sim \dot{M}$ and that the star will be X-ray bright.  We
parameterize the time of transition from the propeller phase to the
tracking phase by $t_{trans}$.

Analyses of accretion onto black holes suggests that the infall will
eventually become an advection-dominated accretion flow (an ADAF;
Narayan \& Yi 1994, 1995; Abramowicz et al. 1995; Narayan, Mahadevan
\& Quataert 1998; Kato, Fukue, \& Mineshige 1998) and that much of the
mass will be ejected prior to reaching the surface of the star
(e.g. Blandford \& Begelman 1999, Menou et al. 1999, Quataert \&
Narayan 1999).  The ADAF transition will occur at a characteristic
time $t\sim t_{ADAF}$, when the associated accretion luminosity falls
to $\approx 0.01 L_E$ (this is a rough estimate, cf.  Narayan \& Yi
1995), where the Eddington luminosity is $L_{E} = 1.8 \times 10^{38}$
ergs/s.  At the beginning of the ADAF phase, we expect the star to be
X-ray bright, but the X-ray luminosity will decline rapidly with time
as $\dot{M_X}$ becomes much smaller than $\dot{M}$.  Thus we expect to
observe bright sources only between $t_{trans}$ and $\sim 2 t_{ADAF}$.
We identify this phase of evolution with the AXPs.  For the parameters
of interest (\S4), the AXP phase corresponds to no more than an order
of magnitude in time, say from $t\sim5000$ yr to $t\sim40000$ yr.  It
is this slender range that we hope to exploit to explain the
surprisingly narrow distribution of AXP periods.

If, on the other hand, the neutron star enters the ADAF phase before
reaching the efficiently accreting tracking phase, then it 
remains a dim ``propeller system'' for most of its lifetime, and is
never seen as a bright AXP.

\section{Details of the Model}

Cannizzo et al. (1990) have shown that once the accretion disk is
established and evolves under the influence of viscous processes, the
accretion rate declines self-similarly according to $\dot{m} \propto
t^{-\alpha}$.  Motivated by these results, we choose to parameterize
the accretion rate in our model by
$$\dot{m}=\dot{m}_{0}, \,\,\, 0<t<T$$
$$\dot{m}=\dot{m}_{0} \biggl( \frac{t}{T} \biggr)^{- \alpha}, \,\,\, t
\geq T, \eqno(1)$$ where $T$ is of order the dynamical time in the
inner parts of the disk early on, and $\dot{m}_{0}$ is a constant,
which we normalize to the total mass of the disk, $M_d= \int_0^\infty
\dot{M_d} dt$, by $\dot{m}_{0} = [(\alpha -1) M_d]/[\alpha
\dot{M}_{E}T]$, assuming $\alpha > 1$.  Cannizzo et al. (1990) find
$\alpha = 19/16$ for a disk in which the opacity is dominated by
electron scattering and $\alpha = 1.25$ for a Kramers opacity.  The
evolution at early times during the initial settling of the disk is
uncertain, and will probably need to be studied numerically.  It is
possible that $\alpha$ could differ from of order unity during an
early phase when the accretion is highly super-Eddington, thereby
compromising the relationship between $M_{d}$ and $\dot{m}_{0}$.
However, our results at late times are insensitive to this initial
transient behavior and the form for the accretion rate given in
equation (1) should be appropriate.

Given the accretion history from equation (1), the time-dependence 
of the stellar spin can be computed from
$$I \dot{\Omega} \, = \dot{J} , \eqno(2)$$ where $I$ is the moment of
inertia of the star and $\dot{J}$ is the torque acting on the star
from the accretion disk.  Throughout, we assume that torque will be
applied only when $R_m < R_{lc}$.  For the range of parameters chosen
below, $R_{m} < R_{*}$ at $t=0$, and hence the accretion disk extends
down to the surface of the star.  Thus, the torque exerted on the star
is $\dot{J} = \dot{M}_{E} R_{*}^{2} \Omega _{K}(R_{*})$ (see, e.g.,
Popham \& Narayan 1991); note that we have assumed that the accretion
rate at the very surface of the star can be at most at the Eddington
rate, with any excess material being blown away before it can reach
the star's surface by the pressure of radiation from the star.  If
$t_{R_{*}}$ is the time at which $R_{m}$ equals $R_{*}$, we take the
subsequent torque to be given by the following simple heuristic
formula (see Menou et. al. 1999):
$$\dot{J} = 2 \dot{M} R_{m}^{2} \Omega _{K}(R_{m}) \Bigl(1 -
\frac{\Omega}{\Omega _{K}(R_{m})}\Bigr) . \eqno(3)$$ When $\Omega$ is
very large compared to the ``equilibrium'' angular velocity $\Omega
_{K}(R_{m})$, we have the propeller phase in which the spin-down
torque is large as a result of the large angular momentum transferred
to the ejected material.  The torque approaches zero as $\Omega$
approaches $ \Omega _{K}(R_{m})$, and becomes a spin-up torque for
$\Omega<\Omega_K(R_m)$ (this never happens in our scenario, except for
a brief period of a month or so very early in the history of the
neutron star, before it enters the propeller phase).

Equation (2) coupled with equation (3) for the torque can be
integrated to yield an analytic solution for $\Omega (t)$ in terms of
incomplete gamma functions for arbitrary $\alpha$.  A particularly
simple form results from the choice $\alpha = 7/6$.  Since this value
is very nearly that found by Cannizzo et al. (1990) for disks
dominated by either electron scattering or Kramers opacities, we will
adopt $\alpha = 7/6$ in what follows.  The solution of equation (3)
for $t > t_{R_{*}}$ is then
$$ \Omega (t) = \Omega (t_{R_{*}}) e^{2 c_{1} (t_{R_{*}}^{1/2} -
t^{1/2})} + 2 c_{2} \bigl[ Ei(2 c_{1} t^{1/2}) - Ei(2 c_{1}
t_{R_{*}}^{1/2}) \bigr] e^{-2 c_{1} t^{1/2}}, \eqno (4) $$ where
$Ei(x) = \int_{- \infty}^{x} (e^{y}/y) dy$ is an exponential integral,
$c_{1} = 2 \dot{m_{0}} \dot{M}_{E} R_{m,0}^{2} T^{1/2} /I$, and $c_{2}
= 2 \dot{m_{0}} \dot{M}_{E} R_{m,0}^{2} \Omega _{K}(R_{m,0}) T /I$.
Quantities subscripted by zero refer to their values at $t=0$.

For arbitrary $\alpha$, the solution to equation (3) in the limit $t
\gg t_{R_{*}}$ can be obtained using an asymptotic expansion, yielding
$\Omega(t) \sim \Omega_{K}(R_{m,0}) (t/T)^{-3\alpha/7}$.  The
corresponding characteristic age, $\tau_c \equiv
-\Omega/2\dot{\Omega}$ and braking index $n\equiv
\Omega\ddot{\Omega}/\dot{\Omega^2}$ then become $\tau_c \sim 7
t/6\alpha$ and $n\sim(7+3\alpha)/3\alpha$.  Note that for $\alpha =
7/6$, $\tau _{c} \sim t$ and $n\sim 3$, which are identical to those
for a radio pulsar spinning down by magnetic dipole radiation.  Thus,
remarkably, the steady timing behavior of a star being spun down by a
fossil accretion disk with $\alpha = 7/6$ will be indistinguishable
from that of a radio pulsar spinning down by emitting magnetic dipole
radiation.

The evolutionary phases described in Section 2 can be identified from
equation (4).  When the first term in equation (4) dominates, $\Omega
\propto e^{-2 c_1 t^{1/2}}$ and spin-down is rapid; this is the
propeller phase.  When the second term dominates, $\Omega \propto
t^{-1/2}$ and the spin period of the star nearly equals the evolving
equilibrium period; this is the tracking phase.  Note that the
spin-down torque is applied by the full $\dot{m}$, which can be vastly
super-Eddington at early times (for the choice of parameters below),
although presumably the X-ray luminosity would never exceed the
Eddington limit.

\section{Representative Results}

For the examples shown here we choose an initial disk mass of
$M_{d}=0.006M_\odot$, an initial dynamical time of $T=1$ ms, an
initial neutron star spin period of $P_0=15$ ms, and $\alpha=7/6$.
Although arbitrary, these are plausible values; for instance, $M_{d}$
is a very small fraction of the likely fallback mass $M_{fb}$, and
$P_0$ is comparable to the estimated initial spin period of the Crab
pulsar.  We assume that neutron stars are born with magnetic field
strengths in the range $\log(B_{12})\sim1-10$ (Narayan \& Ostriker
1990).  Figure 1 shows the spin histories of 10 neutron stars with
field strengths spanning this range.

For the assumed values of $M_{d}$, $T$, $P_0$, and $\alpha$, a neutron
star with a field strength $B_{12}\sim 5-10$ becomes an AXP (Fig. 1).
Early on, for a period ranging from a few days to up to a year, 
depending on the magnetic field strength, the star accretes at the 
Eddington rate and is very bright.  This phase is not visible since 
it is lost in the emission from the supernova explosion.  As $\dot M$ 
decreases with time, $R_m$ increases, and the system switches to a 
much dimmer propeller phase which lasts for about $10^4$ yr.  In this
phase, the star spins too rapidly to accrete much mass; it spins down 
rapidly, however, as a result of the large decelerating torque due to 
the material that is flung out by centrifugal forces.  Ultimately, 
at $t=t_{trans}\sim 10^{4}$ yr, the star achieves quasi-equilibrium 
with the accretion disk and enters the tracking phase where it 
becomes bright in X-rays.  For the examples shown, the luminosity is 
initially $L_x\sim 10^{36.5}~{\rm erg\,s^{-1}}$, with $L_x$ decreasing
as $t^{-\alpha}$.  The bright phase lasts for only a short time.  When
$t=t_{ADAF}\sim19000$ yr (i.e., when $\dot m\sim0.01$), the accretion
flow switches from a thin disk to an ADAF.  Thereafter, the mass
accretion rate onto the neutron star falls rapidly as a result of
heavy mass loss in a wind.  By $t\sim 2\times
t_{ADAF}\sim10^{4.6}$ yr, the mass accretion rate is likely to be
quite low and the system will be dim once again.

A neutron star with an intermediate magnetic field strength, such as
$B_{12}=4$ (Fig. 1), enters the efficiently accreting tracking phase 
at a time $t_{trans}$ which is greater than $2\times t_{ADAF}$, the 
cutoff for visibility assumed in this model. Such a ``propeller 
system'' would thus remain in the dim propeller phase for almost the 
whole of its lifetime before directly entering the ADAF phase.

For smaller field strengths, $B_{12}\sim1-3$, the behavior is quite
different (Fig. 1).  Like their strong field counterparts, these stars
have an initial brief Eddington-bright accretion phase, followed by a
propeller phase.  However, because of their smaller $R_m$ (a
consequence of their smaller $B_{12}$), the spindown due to accretion
is less effective and $R_{lc}$ remains small.  Therefore, with time,
as $\dot M$ is reduced, a stage is reached when $R_m$ becomes greater
than $R_{lc}$.  At this point, accretion ceases completely and the
neutron star becomes a radio pulsar.  The pulsar phase then proceeds
in the normal fashion, lasting for millions of years until the star
crosses the ``death line'' (e.g. Bhattacharya \& van den Heuvel 1991).
By this time, the fossil disk would most likely have dissipated
completely, so it is very unlikely that the dead pulsar would be
resurrected in a late phase of accretion.

Figure 2 exhibits the different phases discussed above in a $B_{12}-P$
diagram.  Note the clean division between radio pulsars and AXPs. Let
$B_{12,RP}$ be the critical field below which neutron stars become
radio pulsars, and let $B_{12,AXP}$ be the critical field above which
neutron stars go through a bright AXP phase. Then, for a choice of
$M_{d}=0.006M_\odot$ and $P_{0}=15$ ms, $B_{12,RP}=3.9$, and 
$B_{12,AXP}=4.2$. The ``propeller'' systems occur for intermediate 
values of field strength. For the above choice of parameters, note
also the narrow range of resulting AXP spin periods.

The precise values of $B_{12}$ that result in radio pulsars and AXPs,
and the period range over which AXPs are seen, both depend on the
parameters of the model, notably $M_{d}$ and $P_{0}$. In the above
example, we have chosen values for these parameters that seem best to
produce results in agreement with observations. While we defer to a
future paper the question of how these conclusions would change were
the parameters to be drawn from broad distributions instead of from
delta functions, we give below the results for a few other choices of
parameter values.

As mentioned above, for $M_{d}=0.006M_\odot$ and $P_{0}=15$ ms,
$B_{12,RP}=3.9$ and $B_{12,AXP}=4.2$; a neutron star with the
critical field $B_{12}=4.2$ would just enter the tracking phase at the
cutoff time $2 \times t_{ADAF} = 38000$ yr with a period of
$7.1$ s; neutron stars with higher fields would enter the tracking
phase before this cutoff time, and thus one would have a distribution
of periods on both sides of 7.1 s.

For $M_{d}=0.006M_\odot$ and $P_{0}=20$ ms,
$B_{12,RP}=0.8$ and $B_{12,AXP}=4.1$; at the cutoff time 
$2 \times t_{ADAF} = 38000$ yr, a neutron star with
$B_{12}=4.1$  would have a period of 6.9 s. For $M_{d}=0.012M_\odot$  
and $P_{0}=15$ ms, $B_{12,RP}=0.5$ and $B_{12,AXP}=2.4$; at the
cutoff time $2 \times t_{ADAF} = 70000$ yr, a neutron 
star with $B_{12}=2.4$ would have a period of 4.3 s.

For $M_{d}=0.006M_\odot$ and $P_{0}=10$ ms, $B_{12,RP}=36.0$ and 
$B_{12,AXP}=4.4$; since $B_{12,RP} > B_{12,AXP}$, any system that does
not become a radio pulsar goes through a bright AXP phase, and there
are no ``propeller systems'' that directly enter the ADAF phase
without going through a brightly accreting phase; a neutron star with
the critical field strength $B_{12}=36.0$ enters the tracking phase at
$t_{trans}=266$ yr with a period of 3.7 s and at the cutoff time $2
\times t_{ADAF} = 38000$ yr has a period of 53.9 s. Similarly, for 
$M_{d}=0.003M_\odot$ and $P_{0}=15$ ms, $B_{12,RP}=31.1$ and 
$B_{12,AXP}=7.4$; a neutron star with $B_{12}=31.1$ enters the tracking
phase at $t_{trans}=713$ yr with a period of 7.3 s and at the cutoff
time $2 \times t_{ADAF} = 21000$ yr has a period of 47.0 s.

Thus, it is clear that not all choices of parameter values produce a
narrow range of periods; it remains to be seen whether this narrow
range would persist if the parameters were drawn from distributions
centered at or near the ``best case'' values used in the figures.

\section{Conclusions}

We have described a model in which a newly-born neutron star can
experience accretion at a time-varying rate from a disk formed as a
result of fallback following a supernova explosion.  As shown above,
for plausible values of the magnetic field strength, initial spin
period, and disk mass ($B_{12} \simgt 5$, $P_{0} = 0.015$ s, $M_{d} =
0.006 M_{\odot}$), a neutron star will be spun down to periods similar
to those of observed anomalous X-ray pulsars, $P\sim 10$ seconds, on
timescales characteristic of the estimated ages of AXPs, $\sim 10,000
- 40,000$ years.  These conditions are met if the object is seen as an
AXP during the tracking phase of evolution discussed previously up to
roughly twice the age at which the accretion flow evolves into an
ADAF, $t_{ADAF}$.  Early during this period, the X-ray luminosity of
the source is high compared with observed AXPs; however, the
luminosity is expected to fall quickly after about $t_{ADAF}$, and so
it could mean, for example, that the known AXPs are seen some time
after $t_{ADAF}$.  For the above parameter values, this model thus
matches the properties of observed AXPs.  Different choices of
parameters (for example, lowering the field strength to $B_{12}=3$)
yields a radio pulsar for an extended period of time. Yet other values
of the surface magnetic field strength (for example, $B_{12}=4$)
produce systems that remain in the X-ray faint propeller phase right 
until the time they make the transition into the dim ADAF phase. 
In this paper, we have merely pointed out the existence of a region 
in parameter space that produces objects with properties resembling AXPs.  
Whether or not this model is capable of accounting for the observed 
relative birthrates of radio pulsars and AXPs will require a more 
thorough analysis.

Our model has a number of advantages over previous theories for AXPs
as accreting sources.  The accretion flow is time-varying, and thereby
provides a natural explanation for the increasing periods of AXPs.
Accretion from a fossil disk evades limits on companion masses
associated with AXPs from timing measurements.  Thompson et al. (1999)
have recently pointed out that large recoil velocities severely limit
the radial extents of disks that would remain bound to a neutron star.
This is not a difficulty for our model, as the disk is tightly bound
initially, before spreading radially as it enters a period of
self-similar evolution.

The other leading model for AXPs invokes ultramagnetized neutron stars
($B\sim 10^{14} - 10^{15}$ G) to account for their timing behavior
(e.g. Thompson \& Duncan 1996) and their X-ray emission (e.g. Heyl \&
Hernquist 1997b, Heyl \& Kulkarni 1998).  An advantage of our scenario
is that it relies on a standard population of isolated neutron stars
having magnetic fields similar to those inferred for radio pulsars and
binary X-ray pulsars and does not require the existence of a separate
class of neutron stars.

How can we discriminate between these two models?  By chance, the
steady spindown caused by accretion from a fossil disk is
indistinguishable from that describing the evolution of ordinary radio
pulsars.  Thus, it appears unlikely that this aspect of AXPs will
unambiguously distinguish these two models.  However, these theories
differ significantly in detail, and many observational tests can, in
principle, separate them.

Torque fluctuations are commonly seen in X-ray binaries (e.g. Bildsten
et al. 1997).  Whether or not the torque fluctuations seen in AXPs can
be explained by our model (or, indeed any of the proposed accretion
scenarios) remains to be seen, but it is at least plausible that an
accretion flow can produce torque noise.  The origin of this behavior
according to the ultramagnetized neutron star model is uncertain.
Proposals include fluctuations induced by glitches (Heyl \& Hernquist
1999) or ``radiative precession'' (Melatos 1999).

The narrow distribution of AXP periods (a factor $\sim 2$) is
puzzling, and lacks a definitive explanation.  In our model, neutron
stars accreting from a debris disk would be visible as AXPs for only a
relatively brief period of time, and hence would be seen in similar
evolutionary states.  Whether or not this is sufficient to reproduce
the observed population of AXPs will depend principally on the
distribution of neutron star-disk systems in the parameter space
defined by magnetic field strength ($B_{12}$), initial spin period
($P_0$), and disk mass ($M_{d}$), and how or if these parameters are
correlated. The other two relevant parameters are $\alpha$ and $T$,
which define the mass accretion rate in equation (1); our long-term
results are quite insensitive to the precise value of $T$, and the
value of $\alpha$ is rather closely fixed by the results of Cannizzo
et al. (1990).

An interesting question concerns the relationship of AXPs to SGRs.
These sources appear similar in most respects, except that SGRs are
subject to occasional, energetic outbursts.  Previously, it has been
suggested that bursts of gamma-ray emission could be produced by
interchange instabilities in the crusts of slowly accreting neutron
stars (Blaes et al. 1990) or by solid bodies impacting neutron stars
(Harwitt \& Salpeter 1973; Tremaine \& Zytkow 1986).  Conceivably,
these processes could play a role in our model, but at present we have
no specific, testable proposal that would distinguish AXPs from SGRs
in this context.

Certainly, the most unambiguous discriminant between the accretion and
ultramagnetized neutron star models would be a direct measurement of
the magnetic field strength (i.e., not derived from timing data).  We
predict that AXPs should have magnetic fields $B\sim 10^{13}$ G, above
average compared with those of radio pulsars, but lying within the
same observed range.  According to the magnetar hypothesis, however,
AXPs have much larger magnetic fields, $B\sim 10^{14} - 10^{15}$ G.
Like some binary X-ray pulsars, it is possible that AXPs will contain
information in their spectra that would provide a determination of the
magnetic field strength.

Existing AXP spectra are well-fitted by thermal profiles with
power-law tails at high energies (e.g.  Corbet et al. 1995,
Oosterbroek et al. 1998, Israel et al. 1999), but lack the resolution
to exhibit discrete spectral features.  Power-law tails at high energy
are characteristic of accreting neutron stars at low luminosities
(e.g. Asai et al. 1996, Zhang, Yu \& Zhang 1998, Campana et al. 1998),
though the high energy tails are typically harder than in the AXPs.
In the ultramagnetized neutron star model, the tails presumably arise
as a result of departures from blackbody emission (e.g. Heyl \&
Hernquist 1998b).  At present, therefore, the spectral information is
ambiguous.  However, the recently deployed Chandra X-ray satellite and
the forthcoming XMM mission promise to revolutionize the field of
high-precision X-ray spectroscopy, and we are thus optimistic that the
true nature of AXPs will be revealed in the near future.

Several arguments against an accretion model of AXPs have been
presented in the literature.  Some of these arguments apply to our
model as well.  Tight limits on optical and especially infrared
counterparts of AXPs (e.g. Coe \& Pightling 1998) severely constrain
the amount of emission that can come from the disk.  For reasonable
parameters, the disks in our model expand to about an AU during the
AXP phase, and it might be difficult to make the disk as dim as the
observations require.  Kaspi, Chakrabarty \& Steinberger (1999) have
shown that the AXP 1E 2259+586 has an extraordinarily low level of
timing noise.  This is hard to explain in an accretion-based model
because the disk, by virtue of its turbulent nature, is likely to
produce a noisy torque.  Note, however, that the fossil disks are
likely to differ from conventional disks in accreting binary systems
by virtue of, e.g., their composition, so the implications of the
observations for our model are uncertain.  Li (1999) has argued that
AXPs cannot respond quickly enough to a changing $\dot M$ to track the
equilibrium spin period of a time-dependent disk.  This is clearly not
a problem for our model.  As Fig. 1 shows, the systems that become
AXPs ($B_{12} \simgt 5$) spin down rapidly enough in the propeller
phase to be able to attain periods close to their (evolving)
equilibrium spin periods in $\sim 10^{3}-10^{4}$ years.

\bigskip
\bigskip

We would like to thank Deepto Chakrabarty, Jeremy Heyl, and Vicki
Kaspi for useful discussion.  This work was supported in part by NSF
grant AST 9820686.

\newpage

\figcaption{The period evolution of a neutron star of initial spin
period $P_{0}=0.015$ seconds and total mass of the surrounding
accretion disk $M_{d}=0.006 M_{\odot}$. The values of $B_{12}$ are 1,
2,...,10, with $B_{12}=1$ for the lowermost curve and $B_{12}=10$ for
the uppermost. Neutron stars with $B_{12} \simlt 3.9$ go into a long
phase as radio pulsars, in which the magnetospheric radius $R_{m}$ is
larger than the light cylinder radius $R_{lc}$. Neutron stars with
$B_{12} \simgt 4.2$ go through a propeller phase of rapid spin-down,
eventually reaching the more slowly evolving tracking phase. The AXP 
phase, as explained in the text, is roughly between $t_{trans}$, the 
time of transition between the propeller and tracking phases, and 
$\sim 2 t_{ADAF}$, where $t_{ADAF}$ is the time at which the accretion 
flow becomes advection-dominated. Stars with intermediate values of 
$B_{12}$ (shown is the case with $B_{12}=4$) remain in the propeller 
stage until they become too dim to be seen at $t\sim 2t_{ADAF}$. 
Although the diagram shows the period evolution for $t > 2 t_{ADAF}$ 
to continue on the tracking phase, the exact evolution at these times 
is uncertain, as it is not clear how the mass outflows during the ADAF 
phase will affect the torque applied by the disk.}

\figcaption{The different possible phases of evolution of a neutron
star of initial spin period $P_{0}=0.015$ seconds and total mass of
the surrounding accretion disk $M_{d}=0.006 M_{\odot}$. All the stars
go through an initial brief stage of Eddington-bright
accretion, during which the star is spun up to the extent of the width
of the bold long-and-short-dashed line on the left. Subsequently, 
stars with $B_{12} \simlt 3.9$ go through a short stage of accretion 
in the propeller phase before becoming radio pulsars during which time 
the magnetospheric radius $R_{m}$ is larger than the light cylinder 
radius $R_{lc}$; they continue in this stage until they reach the 
pulsar death line, given by $B_{12} \approx 0.2 P^{2}$ (Bhattacharya 
\& van den Heuvel 1991). On the other hand, stars with $B_{12} \simgt 
4.2$ go through a dim propeller phase for $\sim 10^{4}$ years, 
eventually reaching the bright AXP phase which, as marked, lasts 
between $t_{trans}$ and $2 t_{ADAF}$; after $t=2 t_{ADAF}$, we assume 
that the AXPs will be too dim to be seen; this region is marked as 
``Dead AXPs''; note that there is no possibility that these latter 
neutron stars will be seen as radio pulsars after the AXP phase, since 
they will be beyond the pulsar death line. Stars with intermediate 
values of $B_{12}$ are ``propeller systems'', which pass directly from 
the propeller phase into the dim ADAF phase. The area hatched with 
solid lines denotes the region in which the neutron star is X-ray 
bright; the area hatched with broken lines denotes the region in which 
the neutron star acts as a radio pulsar.}

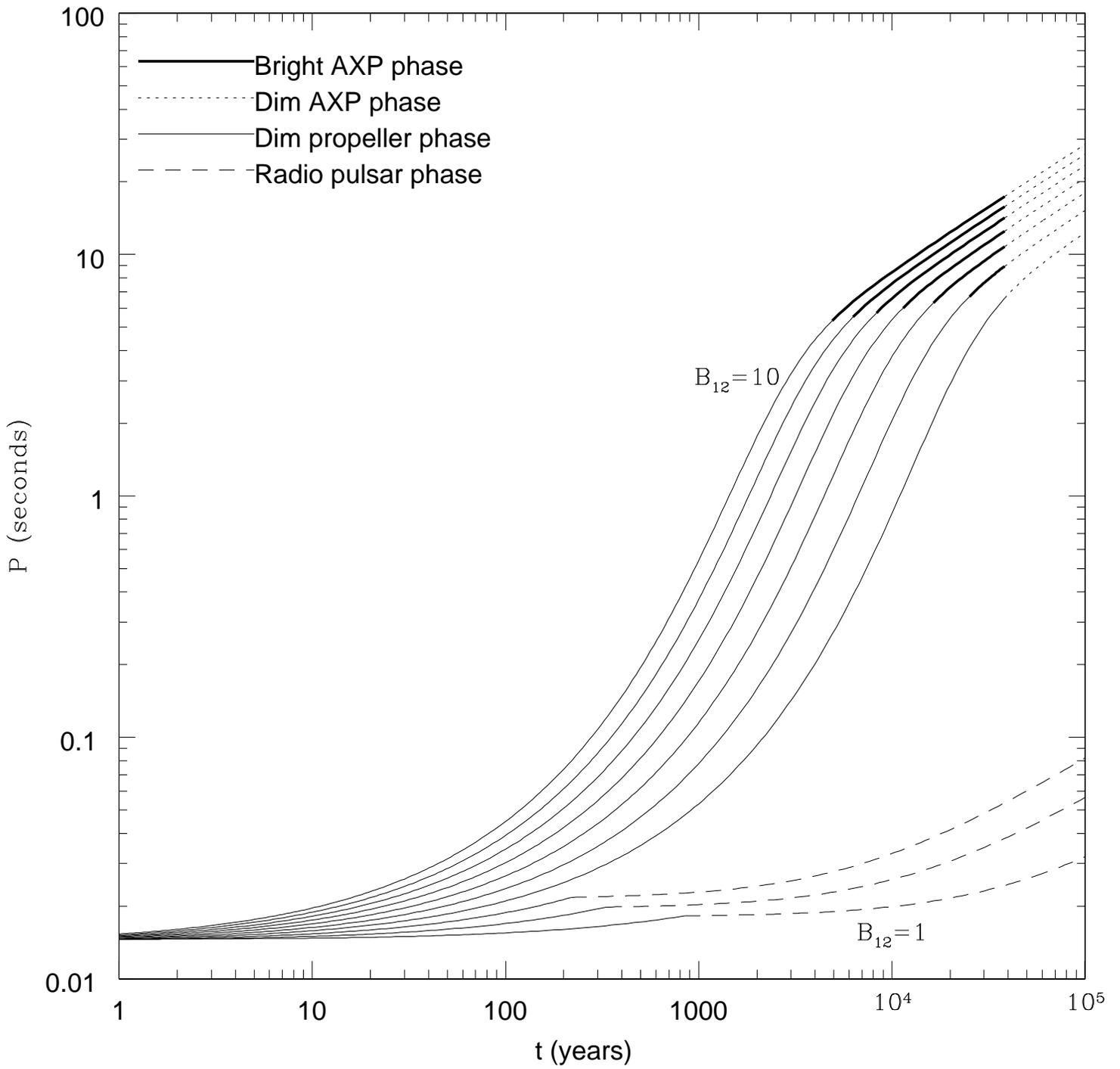
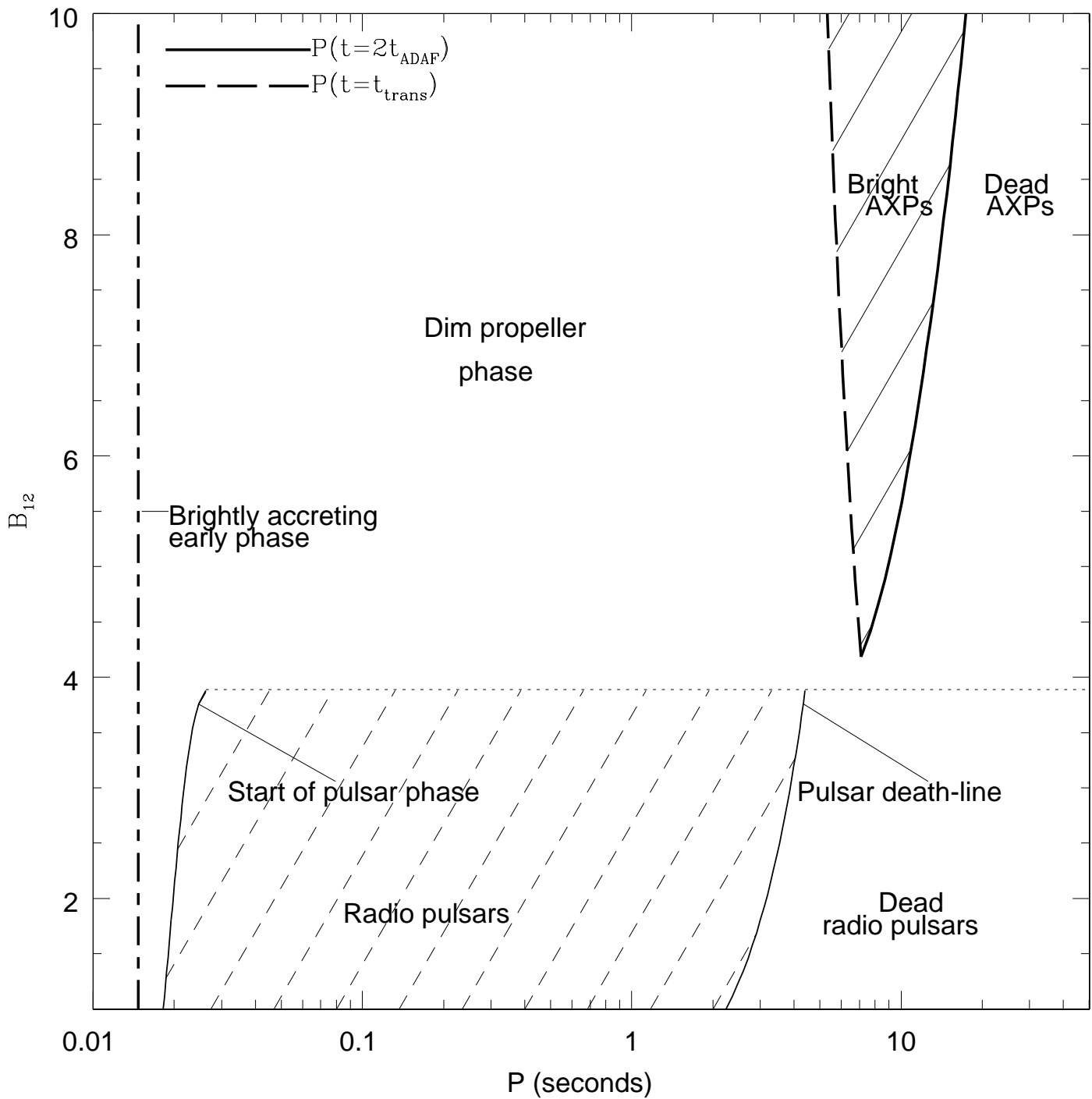
\end{document}